\documentstyle [titlepage,twoside,11pt]{article}
\begin{document}

\def\be{\begin{equation}}
\def\ee{\end{equation}}
\def\nono{{\nonumber}}
\def\tqr{\textstyle\frac14}
\def\call{{\cal L}}
\def\cald{{\cal D}}
\def\calm{{\cal M}}
\def\thf{\textstyle{1\over 2}}
\def\tqr{\textstyle{1\over 4}}
\def\roe{\textstyle{3\over 8}}
\def\tfn{\textstyle{4\over 9}}
\def\tos{\textstyle{1\over 6}}
\def\ttt{\textstyle{2\over 3}}
\def\tot{\textstyle{1\over 3}}
\def\toe{\textstyle{1\over 8}}
\def\ton{\textstyle{1\over 9}}
\def\tfrac{\textstyle\frac}
\def\c@ncel#1#2{\ooalign{$\hfil#1\mkern1mu/\hfil$\crcr$#1#2$}}
\def\slp{\partial\!\!\!/}
\def\slnp{p\!\!\!/ }
\def\slA{A\!\!\!\!/ \,}  
\def\slB{B\!\!\!\!/ \,}  
\def\slW{W\!\!\!\!\!/ \,} 

\begin{titlepage}
\begin{flushright}
UF-IFT-HEP-99-6\\ Original: June 1997\\ Revised: June 1999
\end{flushright}
\ \\
\begin{center}
{\bf ELECTROWEAK THEORY WITHOUT HIGGS BOSONS}
\ \\ \ \\
Angus F.~Nicholson\footnote{Deceased}\\
{\it Department of Theoretical Physics, IAS}\\
{\it The Australian National University, Canberra, ACT 0200, 
Australia}\\
\ \\ \ \\
Dallas C.~Kennedy\footnote{e-mail: kennedy@phys.ufl.edu}\\
{\it Department of Physics, University of Florida,}\\
{\it Gainesville, Florida 32611, USA}\\
\ \\ \ \\ \ \\
{\bf ABSTRACT}
\end{center}

A perturbative $SU(2)_L\times U(1)_Y$ electroweak theory containing $W$, $Z$, 
photon, ghost, lepton and quark fields, but no Higgs or other fields, gives 
masses to $W$, $Z$ and the non-neutrino fermions by means of an unconventional 
choice for the unperturbed Lagrangian and a novel method of renormalisation. 
The renormalisation extends to all orders. The masses emerge on 
renormalisation to one loop. To one loop the neutrinos are massless, the 
$A\leftrightarrow Z$ transition drops out of the theory, the $d$ quark is 
unstable and $S$-matrix elements are independent of the gauge parameter $\xi$.
\ \\ 
\begin{center}
{\bf International Journal of Modern Physics A 15} (2000) 1487
\end{center}

\end{titlepage}

\noindent{\large\bf 1. Introduction}
\medskip

It is widely considered that the Standard Model may be a low-energy effective 
field theory and that the Higgs boson might not exist (see e.g.\ [1,2]). This 
paper describes a perturbative $SU(2)_L\times U(1)_Y$ electroweak theory that 
contains only $W$ and $Z$ bosons, the photon, ghosts and $n$ generations of 
leptons and quarks, and in which $W$, $Z$ and fermions gain masses via 
renormalisation, in a way illustrated by a toy model [3]. There 
are no scalar fields, nor any new particles such as technifermions or preons, 
in the theory. 

The effective Lagrangian density $\call$, given by (6) (section~2), is similar 
to that of standard electroweak (GSW) theory with its Higgs sector omitted; 
also, for simplicity, family mixing has been suppressed. 
The action is $SU(2)_L\times U(1)_Y$ BRS-invariant, as it is in GSW theory. We 
make the decomposition $\call = \call_0 + \call_1$ and renormalise
in an unconventional way. The physical masses of $W$, $Z$ and the fermions 
emerge on renormalisation to one loop. The theory is renormalisable, by 
the method proposed, to all orders.

Renormalising to one loop, we are able to choose values for the counterterm 
parameters that give masses to the $W$ and $Z$ bosons while keeping the photons 
massless, give suitable masses to the fermions, renormalise the vertices to 
their usual forms (with renormalised parameters $g_R$, $\theta_R$ that equal 
the unrenormalised quantities $g$, $\theta$) and satisfy the conditions for 
$SU(2)_L\times U(1)_Y$ invariance. This
renormalisation requires the neutrinos to be massless and $m_W= m_Z \cos\theta$ 
to hold, where $\theta$ is the Weinberg angle transforming from $W_\mu^3$, 
$B_\mu$ to $Z_\mu$, $A_\mu$ fields. The resulting propagators and vertices, and 
so $S$-matrix elements, are independent of the gauge parameter $\xi$ in 
$\call$. Also, the loop quantity $\pi_{\mu\nu}^{AZ}$ that mediates $A 
\leftrightarrow Z$ is finite, with the result that the divergent 
renormalisation factors $Z_3^{1/2}$, $Z_3^{1/2}$ from the contiguous $A$, $Z$ 
lines cause the $A\leftrightarrow Z$ transition to vanish from the theory, at 
least to one loop. Further, we do not need to renormalise the left and right 
components of fermion propagators separately. 

In addition to this one-loop development, we outline, for the $W$, $Z$, photon 
and charged lepton propagators, and (as an example) the $WWZ$ vertex, proposed 
procedures for renormalising the theory to all orders, while continuing to 
fix arbitrary boson and fermion masses.

How should $\call_0$ be chosen? The norm is to take the quadratic part of a 
given $\call$ (prior to the insertion of any $Z_i=1-c_i$ renormalisation 
factors) to be $\call_0$. Thus, for 
\be 
\tilde \call = -{\tqr} (\partial_\mu A_\nu - \partial_\nu A_\mu)^2 + \bar\psi 
(i\slp - e \slA)\psi , 
\ee
traditionally one would take for $\tilde\call_0$ the quadratic part explicitly 
present here; then $\tilde\call$, with a gauge-fixing term added and $Z_i$ 
factors inserted, would define a massless QED theory. However, we recall that 
Schwinger [4] gave an exact solution from (1) in 1 + 1 spacetime dimensions, 
in which the only physical particle is a massive vector boson (see [5]). Also, 
we recall that $\call$ in GSW theory, in its initial form, does not contain 
vector boson mass terms of the 
form $M^2 A_\mu A^\mu$, yet that theory gives massive $W$ and $Z$ bosons. From a
contemporary perspective, in which $\call$ and its symmetries are central, it 
appears to be legitimate, given $\call$, to admit any choice for $\call_0$ that 
leads to a theory that is self-consistent and, in the case of a physical 
theory, fits experiment. As regards gauge symmetry, i.e.\ the BRS symmetry of 
the action, we recall that in QED, QCD and GSW theory, while the action $S=\int 
d^4 x\call$ is invariant under the appropriate BRS transformations, the 
partial-actions $S_{0,1} = \int d^4 x \call_{0,1}$ are not, so that gauge 
invariance places no immediate condition on the choice of $\call_0$; 
ultimately, that choice must lead to a gauge-independent $S$-matrix. As in QED 
and QCD, the gauge symmetry is broken in the present theory in going from 
$\call$ to $\call_0$; however, we find that $S$-matrix 
elements are independent of the value of the gauge parameter $\xi$ in $\call$, 
(6), as is the case in QED, QCD.  

We make a decomposition, $\call= \call_0 + \call_1$, in which $\call_0$ 
contains fermion mass terms. Using the fields defined in section~2, we place 
the mass term
\be
\call_m = - \sum (m_{ej1} \bar e_j e_j +m_{uj1} \bar u_{j\alpha} u_{j\alpha} + 
m_{dj1} \bar d_{j\alpha} d_{j\alpha})  
\ee
in $\call_0$ and place $-\call_m$ in $\call_1$, so that $\call$ is unchanged 
and the $SU(2)_L \times U(1)_Y$ invariance is unbroken. With the usual definition 
of  ``bare", the bare masses of all particles are zero, since there are no 
masses in $\call$. We refer to the masses in (2) as  ``initial" masses. For a 
stable fermion, the initial mass $m_1$ is later put equal to the renormalised 
mass $m_R$. The 
theory does not predict or impose values for the masses of non-neutrino 
fermions (or of $W$, $Z$) and in that sense the masses $m_1$ are arbitrary. 
This arbitrariness may be compared with that of the Yukawa coupling parameters 
that determine the fermion masses in GSW theory. The splitting of zero made to 
place $\call_m$, $-\call_m$ in $\call_0$, $\call_1$ is logically on the same 
footing as the splitting made of $m{\bar\psi}\psi$ in QED, to place $m_{\rm
phys} {\bar\psi}\psi$ in $\call_0$ and $\delta m{\bar\psi}\psi$ in $\call_1$. It
might appear that the $\call_m$, $-\call_m$ step must be nugatory, since 
resulting mass terms from $\call_0$, $\call_1$ cancel in the denominators 
of the full, improper propagators, i.e.\ in $\{ \slnp - m - [-m +\sum(\slnp) 
+ c.t.]\}$; however, the $m_{j1}$ also contribute to the self-energy functions 
$\sum(\slnp)$ and $\pi_{\mu\nu}(k)$ for the fermions and $W$, $Z$ bosons, 
leading to masses for these particles.

The self-energy tensor for $W$ or $Z$ can be written in the form
\be
\pi_{\mu\nu}(k) = (k_\mu k_\nu- g_{\mu\nu}k^2) \pi(k^2) +k_\mu k_\nu \tau(k^2) 
+g_{\mu\nu} \rho(k^2)                              
\ee
with $\rho(k^2)$ defined not to contain a factor $k^2$. The terms 
$\rho_W(k^2)$, $\rho_Z(k^2)$ are central to the generation of the $W$ and $Z$ 
masses in this theory. To one loop, only fermion loops contribute to 
$\rho(k^2)$, and $\tau(k^2)$ is zero. The development requires that the initial 
fermion masses $m_{j1}$ are of $O(g^k)$, $k\ge 1$, and we take $k=1$ in this 
paper. Then it turns out that the $W$ and $Z$ masses are $O(g)$, as they are in 
GSW theory.  

We use dimensional regularisation (which is usual in electroweak theory [6,7]),
working in $d=4-2\epsilon$ dimensions. Much work has been done on the 
problem of $\gamma^5$ in this regularisation and its variant, dimensional 
reduction, and the various approaches have given the same results in a variety 
of calculations to two and three loops (see e.g. [8--14]). Using the 
prescription of a formal $\gamma^5$ that is totally anticommuting with 
$\gamma^\mu$ in $d$ dimensions [15], and the couplings and propagators given in 
section~2, a calculation to one loop of the divergent parts of $\rho_W(k^2)$ 
and $\rho_Z(k^2)$ gives
\be
 \rho_{W\epsilon}^{(2)} = \rho_{Z\epsilon}^{(2)} \cos^2\theta = 
\frac{g^2}{32\pi^2\epsilon} \sum(m_{ej1}^2 + 3 m_{uj1}^2 +3m_{dj1}^2).     
\ee
Since the $SU(2)_L\times U(1)_Y$ invariance does not permit $\call$, (6), to 
contain gauge boson mass counterterms, these terms cannot be directly 
cancelled. In our renormalisation procedure we use the $c_W$, $c_B$ counterterm 
parameters in $\call$ to cancel the most divergent part (at any given order) of 
$k^2 \pi_W (k^2) -\rho_W(k^2)$ at $k^2 = m^2_W$, and similarly for $Z$. 

The $W$ and $Z$ bosons and many of the fermions are unstable; however, the 
usual perturbative and LSZ formalisms [16] do not accommodate unstable 
particles in a consistent way (see e.g.\ [17-19]). For example, in the usual 
approach the initial and final state vectors $|i\rangle,\,|f\rangle$ of a 
matrix element $S_{fi}$ represent physical states that at $t\to\pm\infty$ 
comprise specified sets of physical particles assumed stable.
Consequently, the well-known unitarity argument for the existence of Higgs 
bosons [20], involving $W$ or $Z$ bosons in initial or final states, is open to
question. 
It is often assumed that stable-particle perturbation theory can be 
used with an unstable particle represented by a propagator in which the mass is 
complex, i.e.\ that
the (principal) denominator of the propagator is of the form $k^2 -\calm^2$, 
with $\calm^2$ complex (see e.g.\ [21,22] for the complex mass of $Z$ in GSW 
theory). Then, if we apply the LSZ reduction formula [7] to a diagram 
containing an unstable particle or an external line, the result is zero, 
since the projection operator for the particle, containing a real mass $m$ 
(from $\call_0$), operates on the propagator, containing a complex mass 
$\calm$. Since the zero result is independent of $m$, the reduction formula 
does not restrict $m$, as it does for a stable particle (to be $m_R$). On this 
basis, we take $m_{W1} = m_{Z1}=0$ in this paper, and renormalise to the 
complex masses $\calm_W$, $\calm_Z$. Further, we see that if we deal with 
unstable particles by means of complex-mass propagators, then the usual 
interpretation by means of the Cutkosky rules [24] of the $S$-matrix unitarity 
relation, which, writing $S=1+iT$, is
\be
T_{fi} - T_{if}^* - i\sum T_{nf}^* T_{ni} =0,                         
\ee
is rendered ambiguous, because for a complex-mass unstable 
particle one cannot identify a state carrying a real mass
corresponds to the physical particle. We propose that the physical 
interpretation should be made at the level of the renormalised theory. These 
issues are problems also for the Standard Model. Questions of gauge invariance 
and unitarity with unstable particles, and the complex $Z$ pole mass, are 
discussed in the literature [25--27]. In this paper we use the 
standard perturbative formalism even though most of the particles are unstable, 
as is commonly done for the Standard Model, but it is clear that we cannot 
deal with unitarity in the absence of a consistent unstable-particle extension 
of perturbation theory. Consequently, we do not discuss unitarity further in 
this paper.

Section 2 deals with the Lagrangian, the $SU(2)_L\times U(1)_Y$ BRS-invariance of 
the action, $\call_0$, quantisation and propagators. Section~3 discusses the 
renormalisation to all orders of the $W$ and $Z$ propagators to obtain $W$ and 
$Z$ masses, and of the photon and ghost propagators; then renormalises these 
propagators to one loop explicitly; also, it shows how $A\leftrightarrow Z$
mixing drops out of the theory to one loop. In section~4 the renormalisation of
charged lepton propagators to all orders is discussed, followed 
by the renormalisation of all lepton and quark propagators to one loop, to give 
suitable masses. In section~5 we deal with the vertices. We outline a proposed 
renormalisation of the $WWZ$ vertex to all orders, then complete the 
renormalisation of the theory to one loop, showing that the resulting 
counterterm parameters satisfy the conditions for $SU(2)_L\times U(1)_Y$ 
invariance. We also point out that the theory renormalised to one loop has 
$S$-matrix elements that are independent of the gauge parameter $\xi$. 
Section~6 sums up these themes and their relation to other, similar work.

\bigskip\bigskip\newpage
\noindent{\large\bf 2. Lagrangian, $SU(2)_L\times U(1)_Y$ invariance, $\call_0$, 
propagators}
\medskip

The Lagrangian density is essentially that of standard GSW theory with Higgs 
fields omitted and quark mixing suppressed. It is
\begin{eqnarray}
\call &=&
-{\tqr} f_W W^k_{\mu\nu} W_k^{\mu\nu} - {\tqr} f_B B_{\mu\nu} B^{\mu\nu} - 
f_6(2\xi)^{-1} [(\partial\cdot W)^2 +(\partial\cdot B)^2)] \nono\\
&&+ {\thf} f_1 g\epsilon^{ijk} W^i_{\mu\nu} W_j^\mu W^\nu_k - {\tqr} f_5 g^2 
\epsilon_{ijk} \epsilon^{i\ell n} W^j_\mu W^k_\nu W^\mu_\ell W^\nu_n \nono \\
&&+ \tilde f_3(\partial_\mu \eta^*_k) \partial^\mu \eta_k +\tilde f_1 g 
\epsilon^{ijk} (\partial_\mu \eta_i^*) \eta_j W^\mu_k +\tilde f_3(\partial_\mu 
\eta^*_B) \partial^\mu \eta_B \nono \\
&& + i\sum \left[ \begin{array}{cr}
f_{2\ell L} \bar\ell_L \slp \ell_L + f_{2eR}\bar e_R \slp e_R + 
f_{2qL} \bar q_{\alpha L} \slp q_{\alpha L}\\
+ f_{2uR} \bar u_{\alpha R} \slp u_{\alpha R} + f_{2dR} \bar d_{\alpha 
R} \slp d_{\alpha R}\end{array}\right] \nono \\
&&- \sum\left\{ \begin{array}{clr}
f_{1\ell L} \bar\ell_L ({\thf} g\tau\cdot\slW)\ell_L + (c_{\ell L} -{\thf} g') 
\bar\ell_L \slB \ell_L + (c_{eR} -g') \bar e_R \slB e_R \\
+ f_{1qL}\bar q_{\alpha L} ({\thf} g\tau\cdot \slW) q_{\alpha L} +(c_{qL} + 
{\tos} g') \bar q_{\alpha L} \slB q_{\alpha L}\\
+(c_{uR} +{\ttt}g') \bar u_{\alpha R} \slB u_{\alpha R} + (c_{dR} - {\tot}g') 
\bar d_{\alpha R} \slB d_{\alpha R}\end{array} \right\},             \nono  \\
&&  
\end{eqnarray}
where $g' = g\tan\theta$, $f_i=1-c_i$, with the $c_i$ ($c_W,c_B,\ldots,c_{2\ell 
L},\ldots$) being counterterm parameters of $O(g^2)$ and higher; 
$W^k_{\mu\nu}\equiv \partial_\mu W^k_\nu -\partial_\nu W^k_\mu$ ($k=1,2,3$) and 
similarly for $B_{\mu\nu}$ and for $A_{\mu\nu}$, $Z_{\mu\nu}$ below; and 
$\eta^*_k$, $\eta_k$, $\eta^*_B$, $\eta_B$ are the ghost fields. The sums are 
over $n$ generations of lepton and quark fields 
$\ell_j(\nu_j,e_j),q_{j\alpha}(u_{j\alpha},d_{j\alpha})$, $j=1,\ldots,n$ and 
over the colour index $\alpha$; suppressing $j$, we have written $\ell_L$, 
$q_{\alpha L}$ for the left-handed doublets, $e_R$, $u_{\alpha R}$, $d_{\alpha 
R}$ for the singlets (the fields $\nu_{jR}$ are absent) and $f_{2\ell 
L},f_{2eR},\ldots$ for $f_{2\ell jL},f_{2ejR},\ldots,f_{1\ell jL}$,  etc. In 
$d=4-2\epsilon$ dimensions, $g$ is to be replaced by $g\mu^\epsilon$, with $g$ 
dimensionless and $\mu$ a scale mass, in the usual way; this is left 
implicit in what follows.

It is straightforward to ensure in the usual way [7,28] that we have an 
$SU(2)_L\times U(1)_Y$ gauge theory. We transform to bare fields $W_\mu^{kb} = 
f_W^{1/2} W_\mu^k,\ldots$ \  and parameters $\xi_W^b = f_6^{-1} f_W\xi$, 
$\xi_B^b = f_6^{-1} f_B\xi$, $g_1^b = f_1 f_W^{-3/2} g$, $g_5^b = f_5^{1/2} 
f_W^{-1} g,\ldots$\,. Then the action $S=\int d^4x\call$ is invariant under an 
$SU(2)_L$ BRS transformation [28] of the bare fields, provided that
\be
\frac{f_W}{f_1} = \frac{f_1}{f_5} = \frac{\tilde f_3}{\tilde f_1} = 
\frac{f_{2\ell jL}}{f_{1\ell jL}} = \frac{f_{2qjL}}{f_{1qjL}} ,  
\ee
which ensures that $g_1^b=g_5^b=\cdots$\,. The action is also invariant under a 
suitable $U(1)_Y$ BRS transformation of the fields without the imposition of any 
conditions, in the usual way.

Returning to the form (6), we transform in the standard way to the fields 
$A_\mu = W^3_\mu sin\theta + B_\mu \cos\theta$, $Z_\mu = W_\mu^3\cos\theta 
-B_\mu\sin\theta$, $W^{\pm}_\mu = (1/\sqrt 2)(W^1_\mu \mp i W\mu^2)$, $\eta^{*+} 
= (1/\sqrt2)(\eta_1^* -i\eta_2^*)$, $\eta^- = (1/\sqrt2)(\eta_1+i\eta_2)$, 
$\eta^{*-} =(1/\sqrt2)(\eta_1^* +i\eta_2^*)$, $\eta^{+}= (1/\sqrt2)(\eta_1 
-i\eta_2)$, $\eta^{*A} = \eta_3^*\sin\theta +\eta_B^*\cos\theta$, 
$\eta^A=\eta_3\sin\theta +\eta_B\cos\theta$, $\eta^{*Z} = \eta_3^*\cos\theta - 
\eta_B^*\sin\theta$ and $\eta^Z = \eta_3\cos\theta - \eta_B\sin\theta$. The 
resulting couplings of the $W$, $Z$, $A$, new ghost and fermion fields are the 
same as those in GSW theory [7]. The quadratic boson-ghost part of $\call$ is 
\begin{eqnarray}
\call_Q 
&=& -{\tqr} [f_A A_{\mu\nu} A^{\mu\nu} + f_Z Z_{\mu\nu}Z^{\mu\nu} +2f_W 
W^+_{\mu\nu}W^{\mu\nu}_- ] +{\thf} c_{AZ} A_{\mu\nu}Z^{\mu\nu} \nono \\
&& - f_6(2\xi)^{-1} [(\partial\cdot A)^2 +(\partial\cdot Z)^2 
+2(\partial\cdot W^+)(\partial\cdot W^-)] \nono \\
&&\tilde f_3 [(\partial_\mu \eta^{*+}) \partial^\mu\eta^- 
+(\partial_\mu\eta^{*-})\partial^\mu \eta^+ + 
(\partial_\mu\eta^{*A})\partial^\mu \eta^A +(\partial_\mu 
\eta^{*Z})\partial^\mu \eta^Z],           \nono\\
&&                
\end{eqnarray}
where
\be
f_{A,Z} = f_{W,B} \sin^2\theta + f_{B,W} \cos^2\theta, 
\ee
\be
c_{AZ} = (c_W-c_B) \sin\theta\cos\theta. 
\ee
The counterterm parameters $c_A$, $c_Z$ are defined by $f_A= 1-c_A$, $f_Z = 1-
c_Z$. 

As discussed in section~1, initial quark and charged lepton masses are introduced 
by placing $\call_m$, (2), in $\call_0$. We also place $-\call_m$ in $\call_1$, 
so that $\call$ is unchanged and the  $SU(2)_L\times U(1)_Y$ invariance is not 
broken. In section~4 the masses $m_{ej1}$, $m_{uj1}$, $m_{dj1}$ are taken to be 
$O(g)$ (and we find that the initial neutrino masses must be zero). This leads, 
in section~3, to $W$ and $Z$ masses of $O(g)$, which is the case also in GSW 
theory. Including $\call_m$, we take $\call_0$ to be
 \begin{eqnarray}
\call_0 &=&
-{\tqr} (A_{\mu\nu} A^{\mu\nu} +Z_{\mu\nu}Z^{\mu\nu} +2W^+_{\mu\mu}W_-
^{\mu\nu} ) \nono \\
&&- (2\xi)^{-1} [(\partial\cdot A)^2 +(\partial\cdot Z)^2 +2(\partial\cdot 
W^+)(\partial\cdot W^-)] \nono \\
&&+ (\partial_\mu \eta^{*+}) \partial^\mu\eta^- +(\partial_\mu \eta^{*-} 
\partial^\mu \eta^+ + (\partial_\mu \eta^{*A}) \partial^\mu \eta^A +(\partial 
_\mu \eta^{*Z}) \partial^\mu\eta^Z \nono \\
&&+ \sum[ \bar\ell(i\slp -m_{\ell 1})\ell +\bar 
q_{\alpha}(i\slp -m_{q1})q_\alpha],  
\end{eqnarray}
in which the fermion part has been abbreviated. If all the particles were 
stable, we could proceed by using the usual path integral or canonical 
quantisation method. We assume that perturbation theory can be extended to
encompass unstable particles so that we
obtain from $\call_0$ the propagators that would result in the stable-particle 
case, viz.
\begin{eqnarray}
iD_{\mu\nu}^{W,Z,A}(k)&=& -ik^{-2} [g_{\mu\nu} +(\xi-1) k_\mu k_\nu k^{-2} ], 
\\
i\tilde D(k^2) &=& ik^{-2}, 
\\
i S_F(\slnp) &=& i(\slnp -m_{j1} )^{-1} , 
\end{eqnarray}
for the massless proto-$W$, proto-$Z$, $A$, ghost and the fermion fields.
Because we do not give a physical interpretation of the basis vectors of the 
state space (or protoparticles) generated by $W_\mu$ and $Z_\mu$, we have the 
simplification that the $W_\mu$, $Z_\mu$ part of $\call_0$ can be quantised 
without the imposition of a gauge condition designed to single out physical 
states. For the $A_\mu$ part of $\call_0$ we can proceed in a standard way. The 
usual quantisation procedure gives, from $\call_0$, a set of particle 
states that includes a unique, nondegenerate vacuum, like that of QED, in which 
the VEV of each field is zero. 

With these propagators, and with the fermion generations and fermion-gauge 
boson couplings the same as they are in GSW theory, there are no anomalies in 
the theory. 

\bigskip\bigskip
\noindent{\large\bf 3. Renormalised boson and ghost propagators and masses}
\medskip

For each boson, labelled by $J= W, Z, A$, we sum the series
\be
i\cald_{J\mu\nu} = i D_{J\mu\mu} + i D_{J\mu\sigma} (i\hat\pi_J^{\sigma \rho}) 
i D_{J\rho\nu} +\cdots, 
\ee
where $\hat\pi_J^{\sigma\rho}$ is defined by
\be
\hat\pi_{J\mu\nu} = (k_\mu k_\nu - g_{\mu\nu} k^2)\hat\pi _J(k^2) + 
k_{\mu}k_\nu \hat\tau(k^2) + g_{\mu\nu} \rho(k^2),  
\ee
in which
\be
\hat\pi_J(k^2) = \pi_J (k^2) -c_J, \qquad \hat\tau_J(k^2) = \tau(k^2) +c_6 
\xi^{-1} ,  
\ee
where $\pi_J(k^2)$, $\tau(k^2)$, $\rho(k^2)$ are given by the self-energy 
integrals with $\rho(k^2)$ defined not to contain a factor $k^2$, and $c_J$, 
$c_6$ are counterterm parameters, with $c_A$, $c_Z$ given by (9). In 
$A\leftrightarrow Z$ subdiagrams the $A$, $Z$ lines are linked by 
$i\hat\pi_{\mu\nu}^{AZ}$, with 
\be
\hat\pi_{\mu\nu}^{AZ} = (k_\mu k_\nu - g_{\mu\nu} k^2) \hat\pi _{AZ} (k^2) 
+k_\mu k_\nu \tau_{AZ}(k^2) + g_{\mu\nu} \rho_{AZ}(k^2),  
\ee
in which
\be
\hat\pi_{AZ}(k^2) = \pi_{AZ}(k^2) - c_{AZ} ,  
\ee
with $c_{AZ}$ given by (10). To one loop (see below), $\tau_J(k^2) =0$, the 
divergent parts of $\rho_W$, $\rho_Z$ are given by (4) and $\rho_A(k^2) =0$, 
which might be true to all orders, as it is in QED.

In the usual way, the sum in (15) is given, correctly to any chosen 
$O(g^{2n})$, by
\be
i\cald_{J\mu\nu} = \frac{-i}{(1+\hat\pi_J)k^2-\rho_J} [ g_{\mu\nu} +k_\mu k_\nu 
Q_J (k^2)],   
\ee
where
\be
Q_J(k^2) = \frac{\xi(\hat\pi_J +\hat\tau_J +1) -1} {k^2 -\xi(\rho_J 
+k^2\hat\tau_J)} . 
\ee

We outline a proposed procedure for renormalising the 
$W$ and $Z$ propagators to all orders to give masses to these bosons, and 
renormalising the photon and ghost propagators to all orders. Then we 
renormalise the $W$, $Z$, photon and ghost propagators to one loop. 

We define the $O(g^{2n})$ component $\pi_{2n}(k^2)$ in $\pi(k^2) = \sum 
\pi_{2n}(k^2)$, and, similarly, $\tau_{2n}$ and $\rho_{2n}$. Following the
conventions of calculations beyond one loop in QED and QCD, in 
which the propagators have the forms (12), (13), (14), and the 
dimensions of coupling parameters and the topologies of diagrams are 
essentially the same as in this theory, we can take $\pi_{2n}(k^2)$ 
(generated in the usual way by $n$-loop diagrams not containing counterterm 
insertions plus $(n-j)$-loop diagrams with insertions) in the form
\be
\pi_{2n}(k^2) = g^{2n} [\pi_{2n,n} \epsilon^{-n} +\pi_{2n,n-1}(k^2,m^2_{j1}) 
\epsilon^{-n+1} +\cdots+ \pi_{2n,0} (k^2, m_{j1}^2)],   
\ee
where $\pi_{2n,n}$ is real and independent of $k^2$, $g^2$ and the initial 
masses $m_{j1}$. Since the $m_{j1}$ are $O(g)$, and we encounter 
$\pi_{2n}(k^2=\calm^2)$ with $\calm$ of $O(g)$, any expansions of 
$\pi_{2n,\sigma} (k^2,m^2_{j1})$, $\pi_{2n,\sigma} (\calm^2, m_{j1}^2) $ as 
series in $g^2$ would cause $\pi_{2n}(k^2)$ to contain $g^{2N}$, $N>n$. 
However, we do not need to make such expansions in order to renormalise the 
theory, and it is convenient to refer to $\pi_{2n}(k^2)$, $\pi_{2n}(\calm^2)$ 
as being of $O(E g^{2n},\epsilon^{-n})$, where $Eg^{2n}$ indicates the 
explicit $g^{2n}$ factor in (22) and $\epsilon^{-n}$ is the highest power of 
$\epsilon^{-1}$ that is present. (This factor $\epsilon^{-n}$ is independent of 
any expansion that might be made of $\pi_{2n,\sigma}$ terms as series in $g^2$. 
When we encounter $g^\sigma \pi_{2n}$ terms we shall extend the notation to 
$O(Eg^{2n+\sigma}, \epsilon^{-n})$.) We treat $\tau(k^2)$ in the same way. On 
dimensional grounds and from the structure of the integrals, $\rho(k^2)$ is of 
the form $\sum m^2_{j1} \rho_j(k^2)$, with $m_{j1}=\beta_{j1} g$ (section~4), so that an 
$(n-2)$-loop diagram free of counterterm insertions generates a component 
of $\rho(k^2)$ of the form (22),  with $\rho_{2n,n} = 
\rho_{2n,n}(\beta_{j1}^2)$, of $O(Eg^{2n} \epsilon^{-n})$, real and independent 
of $k^2$ and $g^2$. Consequently, an $n$-loop diagram generates terms 
$\pi_{2n}(k^2)$, $\rho_{2n+2}(k^2)$ of $O(Eg^{2n}, \epsilon^{-n})$, 
$O(Eg^{2n+2},\epsilon^{-n})$. To $N$ loops (summing over appropriate 
lower-order diagrams up to $N$-loop diagrams free of counterterm insertions) we 
have the quantities $\pi_{(N)}(k^2)$, $\rho_{(N)}(k^2)$, of 
$O(Eg^{2N},\epsilon^{-N})$, $O(Eg^{2N+2},\epsilon^{-N})$. Writing 
$(\partial/\partial k^2)\pi (k^2) = \pi'(k^2)$, etc., we see from (22) that
\be
\frac{\pi'_{(N)}(k^2)}{\pi_{(N)}(k^2)} \to 0, \qquad \frac{\pi''_{(N)}(k^2)} 
{\pi_{(N)}(k^2)} \to 0, \quad \ldots   
\ee
as $\epsilon\to 0$, and similarly for $\pi'_{2n}, \pi_{2n}$, and 
$\rho',\rho$, etc. 

It is sufficient for $c_J$ to be of the form
\be
c_J = \sum c_{J,2n} g^{2n} \epsilon^{-n},  
\ee
i.e.\ that the counterterm parameters $c_W$, $c_Z$, $c_A$ need not contain 
components of order $g^{2n}\epsilon^{-\sigma}$, $\sigma < n$. We define, for 
$J=W, Z$,
\be
\hat\pi_{JNL} = \sum_{n=1}^N (\pi_{J,2n,n} - c_{J,2n}) g^{2n} \epsilon^{-n},  
\ee
\be
\rho_{JNL} = \sum_{n=1}^N \rho_{J,2n+2,n} g^{2n+2}\epsilon^{-n},  
\ee
which are independent of $k^2$ and real. Clearly $\hat\pi_{JNL}$, $\rho_{JNL}$ 
are the leading divergent parts of $(\pi_J(k^2) -c_J)$, $\rho_J(k^2)$ taken to 
$N$ loops. They are of $O(g^{2N} \epsilon^{-N})$, $O(g^{2N+2}\epsilon^{-N})$ 
(which we rely on below; they are also $O(Eg^{2N},\epsilon^{-N})$, 
$O(Eg^{2N+2},\epsilon^{-N})$). 

To $N$ loops, and for $J=W,Z$, the principal denominator in $\cald_{J\mu\nu}$, 
(20) is easily seen to be 
\be
d_{JN}(k^2) = (1+\hat\pi _{JNL}) k^2 - \rho_{JNL} + R_{JN}(k^2)_{(2N,-N+1)},  
\ee
where the $(2N,-N+1)$ subscript indicates that $R_{JN}(k^2)$ is of 
$O(Eg^{2N},\epsilon^{-N+1})$. For the derivatives of $d_{JN}(k^2)$ we then have
\be
d'_{JN}(k^2) = 1+\hat\pi_{JNL} + R'_{JN}(k^2)_{(2N,-N+1)},    
\ee
\be
d_{JN}^{(\sigma)}(k^2) = R^{(\sigma)}_{JN}(k^2)_{(2N,-N+1)}, 
\ee
where $(\sigma)$ indicates any derivative beyond the first. 

We renormalise the $W$ and $Z$ propagators to have the (principal) pole masses 
$\calm_W$, $\calm_Z$, with 
\be
\calm^2_J = m^2_J - i\delta_J \quad(J=W,Z),   
\ee
where
\be
m_J = \beta_J g, \quad \delta_J = \delta_{J4}g^4 +\delta_{J6}g^6+\cdots  
\ee
and $m_J$, $\delta_J$ are arbitrary (to be fitted to experiment in a complete 
physical theory). (As usual, the renormalisation of the $W$, $Z$ and photon 
propagators is to proceed, order by order, in step with the order by order 
renormalisation of ghost and fermion propagators and vertices as discussed below, 
to fix, at each order, the counterterms needed for renormalisation at
higher orders.) We 
write
\be
d_{JN}(k^2) = d_{JN}(\calm^2) +d'_{JN}(\calm^2) (k^2-\calm_J^2) +\frac1{2!} 
d''_{JN}(\calm^2)(k^2-\calm_J^2)^2 +\cdots\ .  
\ee
From (27) we obtain
\begin{eqnarray}
d_{JN}(\calm^2)
& = (1+\hat\pi_{JNL}) m_J^2 -[1+\hat\pi_{J(N-1)L}]i\delta 
+ O(Eg^{2N+4},\epsilon^{-N}) \nono \\
&\qquad\qquad - \rho_{JNL} +R_{JN}(k^2)_{(2N,-N+1)}. 
\end{eqnarray}

We impose the mass condition
\be
\hat\pi_{JNL} m^2_J - \rho_{JNL} = 0 
\ee
at $N=1,2,\ldots$ loops; i.e.\ at every order $n$ we take, using (25), (26), 
(31),
\be
c_{J,2n} = \pi_{J,2n,n} - \beta_J^{-2} \rho_{J,2n+2,n}.  
\ee
At each $O(g^{2n})$ we are cancelling the most divergent $(\epsilon^{-n})$ part 
of the $g_{\mu\nu}[-k^2\pi(k^2) +\rho(k^2)]$ component of $\pi_{\mu\nu}(k)$, 
(3), at $k^2=m_J^2$. We may compare this with the usual subtraction procedures 
in QED and QCD, in which (with $\rho(k^2) =0$) the divergences in $\pi$ are 
fully cancelled, order by order, at every $O(g^{2n}\epsilon^{-\sigma})$, 
$\sigma\le n$. In this theory it is not necessary to cancel all the divergent parts 
of the $g_{\mu\nu}$ component of $\pi_{\mu\nu}$, as is shown in what 
follows. 

We define
\be
\widehat Z_{JN} = (1+\hat\pi_{JNL})^{-1},  
\ee
which turns out to be the renormalisation factor. We see from (33), (28), (29), 
using (30), (31), (25), (26) and (34), that, correctly to $O(g^{2N+2})$ 
($\hat\pi_{JNL}$, $\rho_{JNL}$ are precisely $O(g^{2N})$, $O(g^{2N+2})$ as well 
as $O(Eg^{2N})$, $O(Eg^{2n+2})$), 
\be
\widehat Z_{JN} d_{JN}(\calm^2) = O(\epsilon),  
\ee
\be
\widehat Z_{JN} d'_{JN} (\calm^2) = 1+ O(\epsilon) ,  
\ee
and
\be
\widehat Z_{JN} d^{(\sigma)}_{JN} (\calm^2) = O(\epsilon).  
\ee
Then from (32) we obtain, correctly to $O(g^{2N+2})$, 
\be
\lim_{\epsilon\to 0} [\widehat Z_{JN} d_{JN}(k^2)] = k^2 -\calm^2. 
\ee
Since $N$ is arbitrary, this result extends to any order. Then
from (20) we see that the renormalised propagator is
\be
i\cald^R_{J\mu\nu} = \frac{-i}{k^2-\calm_J^2} \{ g_{\mu\nu} +k_\mu k_\nu 
\lim_{\epsilon\to 0} [Q_J(k^2)]\}   
\ee
for $J=W,Z$ (we deal with $Q_J(k^2)$ below), and that the renormalisation 
factor is, to $N$ loops, $\widehat Z_{JN}$.  

For the photon we have the denominator
\be
d_A(k^2) = [1+\pi_A(k^2) -c_A] k^2 - \rho_A(k^2).  
\ee
We define
\be
\widehat Z_{AN} = (1+\hat\pi_{ANL}) ^{-1} ,   
\ee
where $\hat\pi_{ANL}$, independent of $k^2$ and real, is defined by (25) with 
$J=A$, while $c_A$ is determined, by (9), from $c_W$, $c_Z$ given by (35); it 
is of the form of $c_J$, (24). To one loop, (55), (63) below show that 
$\rho_A(k^2) =0$ and that $\hat\pi_A$ diverges. If $\rho_A(k^2) =0$ to all 
orders, as holds true in QED, or if $\rho_A$ is less divergent than $\hat\pi_A$ 
at each order, then the photon remains massless. If that is so, we have
\be
d_{AN}(k^2) = (1+\hat\pi_{ANL}) k^2 - \rho_{AN} (k^2) _{(2N+2,-N+\sigma)} + 
O(Eg^{2N},\epsilon^{-N+1})  
\ee
and, correctly to $O(g^{2N})$, obtain 
\be
\lim_{\epsilon\to 0} [\widehat Z_{AN} d_{AN} (k^2)  ] = k^2  
\ee
for arbitrary $N$, so that $\widehat Z_{AN}$ is the renormalisation factor and 
(we deal with $Q_A(k^2)$ below) the renormalised propagator is
\be
i\cald^R_{A\mu\nu} = \frac{-i}{k^2+i\epsilon'} \{ g_{\mu\nu} +k_\mu k_\nu 
\lim_{\epsilon\to 0} [Q_A(k^2)]\}.   
\ee

To determine $Q_W(k^2)$ we could choose $c_6$, order by order, such that, to 
any given $O(g^{2N})$, 
\be
\lim_{\epsilon\to 0} [\hat\tau_{WNL} (\hat\pi_{WNL})^{-1}] = -\lambda^{-1}  
\ee
where $\hat\tau_{WNL}$ is defined analogously to $\hat\pi_{WNL}$ (see (17), 
(25)) and $\lambda$ is arbitrary real.  Then, using (21), (34), it is easy to 
show that (41), for $J=W$, becomes
\be
i\cald^R_{W\mu\nu} = \frac{-i}{k^2-\calm_W^2 +i\epsilon'} \left[ g_{\mu\nu} 
+(\lambda-1) \frac{k_\mu k_\nu}{k^2 -\lambda m_W^2+i\epsilon'} \right], 
\ee
which has the form of the $W$ propagator in GSW theory [7] except that the 
gauge parameter $\xi$ has been replaced by $\lambda$, arbitrary and independent 
of $\xi$ (also, $m_W$ in the principal denominator has been replaced by the 
renormalised mass $\calm_W$). To one loop, it follows from the results given 
below that, with this choice of $c_6$, the $Z$ and $A$ propagators would also 
be of this form, with $\calm_W$, $m_W$ replaced by $\calm_Z$,  $m_Z$; 0, 0.

However, we choose $\lambda = 0$; more correctly, we do not impose (47) 
but instead take $c_6$ to be
\be
c_6 = \sum c_{6(2n)} g^{2n} \epsilon^{-n-1},  \quad c_{6(2n)} \ne 0.  
\ee
Then (21) gives
\[
\lim_{\epsilon\to 0} [Q_{W,Z,A} (k^2)] = -k^2 
\]
and, to all orders, the $W$, $Z$ propagators (41) become
\be
i\cald^R_{J\mu\nu}(k) = \frac{-i}{k^2 - \calm^2_J +i\epsilon'} \left[g_{\mu\nu} - 
\frac{k_\mu k_\nu}{k^2+i\epsilon'} \right] .  
\ee
If $\rho_A$ is zero or less divergent than $\hat\pi_A$ at each order, then the 
renormalised photon propagator is also of the form (50), with $\calm_J$ 
replaced by zero. We see below that this is so to one loop.  These 
Landau-gauge-like propagators are independent of the arbitrary gauge parameter 
$\xi$.

For each ghost, the self-energy integrals are similar to those of QCD. The 
general self-energy integral contains the factor $(k+p)_\mu k_\nu$ coming 
from the terminal vertices, where $k$, $p$ are the ghost and a loop momentum. 
The self-energy is $i\tilde\pi = i k^2 \tilde\pi_0(k^2)$, the same for every 
ghost $(\eta^+, \eta^-, \eta^Z, \eta^A)$. Summing the usual series gives, for 
each ghost, the renormalised propagator 
\be
i\cald^R_\eta (k^2) = -i[k^2 +i\epsilon']^{-1}  
\ee
and the renormalisation factor
\be
\tilde Z_3 = (1+\tilde\pi_0 -\tilde c_3)^{-1}.   
\ee

It has been shown by 't Hooft [29,30] that the contributions to $S$-matrix 
elements made by the poles at $k^2=0$ in the gauge boson and ghost propagators 
(50), (51) cancel to zero. This
is the same cancellation that occurs in GSW theory between the contributions 
from the poles at $k^2 = \xi m_W^2$, $k^2 = \xi m^2$ in the $W$, $Z$ and ghost 
propagators.

It is possible that the renormalised 
propagators should be identified with physical bosons and 
fermions. However, such an interpretation would require the putative 
unstable-particle extension of perturbation theory suggested in 
section~1, and we do not pursue this question here.

We now restrict the discussion to one loop, discarding unnecessary indices. 
Using the propagators (12), (13), (14) we calculate the values
\be
\tau_W = \tau_Z = \tau_A = \tau_{AZ} =0   
\ee
so that, by (17),
\be
\hat\tau_W = \hat\tau_Z = \hat\tau_A = c_6\xi^{-1},  
\ee
and
\be
\rho_A = \rho_{AZ} =0,  
\ee
and the divergent parts
\begin{eqnarray}
\pi_{W\epsilon} &=& \omega (\xi-\tfrac{13}{3}) +\tfrac43 \omega n,   \\  
\pi_{Z\epsilon} &=& \omega (\xi-\tfrac{13}{3})\cos^2\theta +\tfrac43 \omega n \sec^2\theta(1-
2\sin^2\theta +\tfrac83\sin^4\theta),    \\
\pi_{A\epsilon} &=& \omega (\xi-\tfrac{13}{3})\sin^2\theta +\tfrac{32}9 \omega n \sin^2\theta,   \\
\pi_{AZ\epsilon} &=& \omega (\xi-\tfrac{13}{3})\sin\theta\cos\theta +\tfrac43 
\omega n\tan\theta(1-\tfrac83\sin^2\theta),   \\
\tilde\pi_{\epsilon} &=& k^2 \tilde\pi_{0\epsilon} = {\thf}\omega (\xi-{3})k^2,    
\end{eqnarray}
together with (4) for the divergent parts $\rho_{W\epsilon}$, 
$\rho_{Z\epsilon}$, where $\omega=g^2(16\pi^2\epsilon)^{-1}$ and $n$ is the number 
of generations. Then $c_W,c_Z$ 
are given by (35), so that
\be
c_W = \omega [\xi +{\tot}(4n-3)] -\frac{\omega }{2m_W^2} \sum(m^2_{ej1} + 3m^2_{uj1} +3 
m^2_{dj1}),  
\ee
and $c_A$, $c_B$, $c_{AZ}$ are given by (9), (10). 

Defining $\theta$ by
\be
m_W = m_Z\cos\theta,   
\ee
it follows from (17) and the results above that
\begin{eqnarray}
\hat\pi_{A\epsilon} &=& \hat\pi_{Z\epsilon} = \hat\pi_{W\epsilon} = 
\rho_{W\epsilon} m_W^{-2} = \rho_{Z\epsilon} m_Z^{-2} \nono \\
&\equiv \hat\pi_\epsilon  
\end{eqnarray}
and that
\be
\hat\pi_{AZ\epsilon} =0 .  
\ee
These results, and so (62), are needed for the complete renormalisation to one 
loop of the whole theory to succeed with the BRS-invariance conditions (7) 
satisfied. 

With the choice (49), which at one loop is $c_6 = c_{6(2)} g^2 \epsilon^{-2}$, 
we obtain the renormalised propagator (50) for the photon (with $\calm_J$ 
replaced by zero) as well as for $W$ and $Z$. If we had imposed (47), then 
(54), (63) would cause (47) to hold also for $Z$ and $A$, and the $Z$ and $A$ 
propagators would take the form of the $W$ propagators (48), with appropriate 
mass replacements. 

We see from (36), (43) and (63) that $W$, $Z$ and $A$ lines carry the common 
renormalisation factor 
\be
\widehat Z_3 = (1+\hat\pi_\epsilon)^{-1}.  
\ee

On renormalising $A\leftrightarrow Z$, there are divergent $\widehat Z_3^{1/2}$
factors from the $A$,$Z$ lines multiplying the insertion $i\pi_{\mu\nu}^{AZ}$, 
which to one loop is finite, by (18), (64), to give zero as $\epsilon\to 0$. 
This simplifies diagrams and the physical interpretation of the $A$ and $Z$ 
propagators. In GSW theory, however, there is a double-pole $A-Z$ propagator 
[21].

\bigskip\bigskip
\noindent{\large\bf 4. Fermion propagators and masses}
\medskip

For each fermion the full propagator is given by
\be
i S'_F(\slnp) = i(\slnp-m)^{-1} (1+\sigma+\sigma^2+\cdots) 
\ee
where
\be
\sigma = \left[\Sigma(\slnp) -m +\kappa \slnp\right] (\slnp -m)^{-1}, 
\ee
in which $m$ stands for the initial mass $m_1$ ($m_{ej1}$, $m_{uj1}$, 
$m_{dj1}$, zero for a neutrino), the first $m$ in $\sigma$ is from $-\call_m$ 
in $\call_1$, and $\kappa = \kappa_1 +\kappa_5\gamma^5$, with, from (6),
\be
\kappa_1 = {\thf} (c_{2\ell jL} +c_{2ejR}), \qquad
\kappa_5 = {\thf} (c_{2\ell jL} -c_{2ejR}) 
\ee
for a charged lepton, or similar expressions for other fermions, 
while $\sum(\slnp )$ (from the self-energy $-i\sum(\slnp)$) is of the form
\be
\Sigma(\slnp) = -a(p^2) \slnp +b(p^2). 
\ee
We write $a(p^2) = a_1(p^2) +a_5(p^2) \gamma^5$, $b(p^2) = b_1(p^2) 
+b_5(p^2)\gamma^5$. 

To sum the series (66) in the usual way with $-m$ in the numerator 
of $\sigma$, we must take each $m$ to be $O(g^k)$, $k>0$. We make the choice
\be
m_{ej1} = \beta_{ej1}g, \quad m_{uj1} = \beta_{uj1}g, \quad 
m_{dj1} = \beta_{dj1}g,   
\ee
 which we write generically as $m=m_1 = \beta_1 g$. Then, correctly to any 
given $O(g^n)$, (66) sums to
\begin{eqnarray}
iS'_F(\slnp)  
&=& i(\slnp -m)^{-1}  (1-\sigma)^{-1} \nono \\
&=& i[\slnp-m-(\Sigma-m +\kappa \slnp)]^{-1}  \\ 
&=& i\{ [1+a(p^2) -\kappa]\slnp -b(p^2) \}^{-1}, 
\end{eqnarray}
using $A^{-1} B^{-1} = (BA)^{-1}$ and (67), (69). In (71) we see the 
cancellation of the $m$ terms from $\call_0$, $\call_1$; however, the presence 
of $m$ in the propagators $-i(\slnp -m_{j1})$ is responsible for the 
generation of the mass term $b(p^2)$, from which we obtain the renormalised 
fermion mass.

We outline the procedure proposed for the renormalisation 
of the charged lepton propagators to all orders. Then we renormalise all the 
fermion propagators to one loop. 

We renormalise each charged lepton propagator to have the pole mass 
\begin{eqnarray}
\calm &=& m_R - i\delta\\   
&=& \beta_{R}g -i(\delta_2 g^2 +\delta_3 g^3 +\cdots),  
\end{eqnarray}
where $m_R$, $\delta$ are arbitrary (to be fitted to experiment) 
and the generation index $j$ has been suppressed. For the electron, 
which is stable, the $S$-matrix reduction formula requires the renormalised and 
initial masses to be equal, i.e.
\be
\calm_{eR} = m_{eR} \ (=m_{e1R}) \ = m_{e1} \ (=m_{e11}).  
\ee
For each charged lepton we write
\begin{eqnarray}
a(p^2) 
&=& a(\calm^2) + a'(\calm^2)(p^2 -\calm^2) + {\thf}! a''(\calm^2) (p^2 
-\calm^2)^2 +\cdots  \nono \\
&=& a(\calm^2) +2\calm a' (\calm^2) (\slnp -\calm) +U_a(\slnp) (\slnp 
-\calm^2)  
\end{eqnarray}
and similarly for $b(p^2)$, in which $U_a(\slnp)$, $U_b(\slnp)$ depend on 
observing the order of factors shown. The denominator in (72) is then
\begin{eqnarray}
d(\slnp) 
&=& [1+a(\calm^2) -\kappa] \calm - b(\calm^2) \nono \\
&&+ [1+a(\calm^2) -\kappa +2\calm^2 a'(\calm^2) -2\calm b'(\calm^2)] 
(\slnp -\calm) \nono \\
&&+ [2\calm a'(\calm^2) + \calm U_a(\slnp) - U_b(\slnp)] (\slnp-
\calm)^2.  
\end{eqnarray}

We make assumptions about the structure of $a(p^2)$, $b(p^2)$ similar to those 
made for $\pi$, $\tau$, $\rho$ in section~3. Writing $b(p^2) = \sum m_{j1} 
b_j(p^2)$ and $a(p^2) = \sum a_{2n}(p^2)$, $b_j(p^2) = \sum b_{j(2n)}(p^2)$, we 
assume that each of $a_{2n}(p^2)$, $b_{j(2n)}(p^2)$ is of the form (22) (with 
$k^2$ replaced by $p^2$) and that $a_{2n,n}$, $b_{j(2n,n)}$ are real and 
independent of $p^2$ and $g^2$. We denote $a(p^2)$, $a(\calm^2)$, $b(p^2)$, 
$b_j(p^2)$, \dots taken to $N$ loops as $a_{(N)}(p^2)$, 
$a_{(N)}(\calm^2)$,\dots, and, following the treatment of $\pi$, $\tau$, $\rho$ 
in section~3, refer to $a_{(N)}$, $b_{j(N)}$, $b_{(N)}$ as being of 
$O(Eg^{2N},\epsilon^{-N})$, $O(Eg^{2N},\epsilon^{-N})$, 
$O(Eg^{2N+1},\epsilon^{-N})$. Equation (22) holds for $a_{(N)}(p^2)$, 
$b_{(N)}(p^2)$. 

It is sufficient for $\kappa$ to have the form
\begin{eqnarray}
\kappa
&=& \sum \kappa_{2n} g^{2n} \epsilon^{-n}  \\ 
&=& \sum (\kappa_{1(2n)} + \kappa_{5(2n)} \gamma^5) g^{2n} \epsilon^{-n}.  
\end{eqnarray}
We define the leading divergent parts of $[a(p^2) -\kappa]$, $b(p^2)$ taken to 
$N$ loops as
\begin{eqnarray}
\hat a_{NL} &=& a_{NL} - \kappa_N \nono \\
&=& \sum_{n=1}^N  (a_{2n,n} -\kappa_{2n}) g^{2n} \epsilon^{-n} ,  \\ 
b_{NL} &=& \sum_{n=1}^N b_{2n+1,n} g^{2n+1} \epsilon^{-n} ,  
\end{eqnarray}
which are independent of $p^2$ and real, in general contain $\gamma^5$ parts, 
and are $O(g^{2n}\epsilon^{-n})$, $O(g^{2n+1}\epsilon^{-n})$ (as well as being 
$O(Eg^{2n},\epsilon^{-n})$, $O(Eg^{2n+1},\epsilon^{-n})$). To $N$ loops we have
\begin{eqnarray}
[a(p^2) -\kappa]_{(N)} &=&  \hat a_{NL} + A(k^2) _{(2N,-N+1)},  \\ 
b(k^2)_{(N)} &=& b_{NL} +B(k^2) _{(2N+1, -N+1)},  
\end{eqnarray}
where the suffices on $A$, $B$ indicate the orders $O(Eg^{2N},\epsilon^{-
N+1})$, $O(Eg^{2N+1},\newline
 \epsilon^{-N+1})$; so that $a'(k^2)$,$a''(k^2)$,
\dots and $b'(k^2),b''(k^2),\ldots$ are of these orders, respectively. We also 
define
\be
\widehat Z_N = [1+\hat a_{NL} ]^{-1} .  
\ee

To $N$ loops, $d(\slnp)$, (77), becomes, using (73), (74), (80), (81), (82), 
(83),
\begin{eqnarray}
d(\slnp)_{(N)} 
&=& (1+\hat a_{NL}) m_R - (1+\hat a_{(N-1)L})i\delta + O(Eg^{2N+2},\epsilon^{-
N}) +O(\epsilon^{-N+1})\nono\\
&&-b_{NL} + [1+\hat a_{NL} +O(\epsilon^{-N+1})] (\slnp-\calm) + [O(\epsilon^{-
N+1})] (\slnp-\calm)^2, \nono\\
&& 
\end{eqnarray}
where the $O(\epsilon^{-N+1})$ terms are of maximum order $(N-1)$ in 
$\epsilon^{-1}$. We impose the mass condition
\be
\hat a_{NL} m_R - b_{NL} =0  
\ee
for $N= 1,2,\ldots$ loops. From (80), (81), (78) it then follows that
\be
\kappa_{2n} = a_{2n,n} - m_R^{-1} b_{2n,n}. 
\ee
At one loop $b(p^2) = m_1 B(p^2)$, i.e.\ $b_{2,1} = m_1 B_{2,1}$, so that (87) 
is $\kappa_2 = a_{2,1} -m_1 m_R^{-1} B_{2,1}$. 

 From (72), the propagator is, to $N$ loops,
\begin{eqnarray}
i S_{F(N)} (\slnp) &=& i [\widehat Z_N^{-1} \widehat Z_N d(\slnp)_{(N)}]^{-1} \nono \\
&=& \frac{i}{\widehat Z_N d(\slnp)_{(N)}}. \widehat Z_N   
\end{eqnarray}
on using $(AB)^{-1} = B^{-1} A^{-1}$. From (74), (84), (85), (86), we see 
that, correctly to $O(g^{2N+1})$, 
\be
\lim_{\epsilon\to 0} [ \widehat Z_N d(\slnp)_{(N)} ] = \slnp -\calm, 
\ee
so that, using (88) and taking the appropriate charged lepton values for 
$a_{2n,n}$, $b_{2n,n}$, the renormalised charged lepton propagator is
\be
iS_F^R(\slnp) = i(\slnp -\calm )^{-1}  
\ee
to all orders, since $N$ is arbitrary. From (88) we see that this propagator is 
multiplied on the right by the renormalisation factor $\hat Z_N$. 

We now restrict the discussion to one loop and discard unnecessary indices 
from $a$, $b$, $\kappa$. We calculate in $d=4-2\epsilon$ dimensions, using the 
propagators (12), (14), and retain only the divergent parts (writing 
$a_\epsilon = a$, $b_\epsilon = b$, $\kappa_\epsilon = \kappa$) since finite 
parts do not contribute, as we saw above. 

For a charged lepton we obtain
\begin{eqnarray}
a &=& a_1 + a_5 \gamma^5 \nono \\
&=& {\toe} \omega \xi[ 3+5t^2 +3(1-t^2) \gamma^5 ], \\ 
b &=& b_1 + b_5 \gamma^5 \nono \\
&=& {\thf} \omega  (\xi +3) t^2 m_1,    
\end{eqnarray}
where $\omega =g^2(16\pi^2\epsilon)^{-1}$, $t=\tan\theta$ and $m_1 = m_{ej1}$. We see 
that $b_5=0$. Using (79), (87) and (84) we obtain
\begin{eqnarray}
\kappa_1 &=& {\toe} \omega \xi(3+5t^2) - {\thf} \left(\frac{m_1}{m_R}\right)_{ej} 
\omega (\xi+3) t^2 , \\ 
\kappa_5 &=& {\roe} \omega  \xi (1-t^2) , \\ 
\widehat Z_{ej} &=& 1-{\thf} \left( \frac{m_1}{m_R}\right)_{ej} \omega  (\xi+3) t^2 
\end{eqnarray}
to $O(g^2)$, and, from (68),
\begin{eqnarray}
c_{2\ell jL} &=& {\tqr} \omega \xi (3+t^2) -{\thf} \left(\frac{m_1}{m_R}\right)_{ej} 
\omega (\xi+3) t^2 , \\
c_{2e jR} &=& \omega \xi t^2 -{\thf} \left(\frac{m_1}{m_R}\right)_{ej} 
\omega (\xi+3) t^2 , 
\end{eqnarray}
where $ej$ labels the charged lepton in generation $j$. For the electron, which is 
stable, we must have $m_1 = m_R$. It appears that we could take $m_1 = m_R$ 
also for the higher charged leptons ($\mu,\tau,\ldots$), which are unstable. 

The renormalisation of each neutrino propagator proceeds similarly. To one 
loop we find that $b_\nu=0$, so that each neutrino must be massless. If we had 
placed nonzero initial neutrino mass terms $m_{\nu j1}\bar\nu_j\nu_j$ in 
$\call_m$, (6), then $b_\nu$ would remain zero, neutrinos would be massless, 
$m_{\nu j}=0$; and then stable because of the nonzero masses of charged leptons, so 
that the $S$-matrix reduction formula would require $m_{\nu j}=m_{\nu j1}$, 
contradicting $m_{\nu j1}\ne 0$. At least to one loop, the theory only 
admits massless neutrinos. To one loop we obtain
\be
a_{\nu} = {\toe}\omega \xi(3+t^2)(1+\gamma^5), 
\ee
and since $\nu_{jR}$ does not appear in $\call$, 
\be
\kappa_{\nu 1} = \kappa_{\nu 5} = {\thf} c_{2\ell jL}  
\ee
with $c_{2\ell jL}$ given by (96). From (98), (99)   we find that, to $O(g^2)$, 
the renormalisation factor is
\begin{eqnarray}
\widehat Z_{\nu j} 
&=& [1+a_\nu -\kappa_\nu]^{-1} \quad\nono \\
&=& 1-{\tqr}\left(\frac{m_1}{m_R}\right)_{ej} \omega t^2 (1+\gamma^5);  
\end{eqnarray}
however, since $\widehat Z_{\nu j}$ stands to the right of the neutrino 
propagator and every neutrino vertex contains $\gamma^\mu(1-\gamma^5)$, this 
factor in effect is equal to $\widehat Z_{ej}$, so that to one loop we have 
the common lepton renormalisation factor in each generation
\be
\widehat Z_{2\ell j} = 1-\Delta_{\ell j}, \qquad
\Delta_{\ell j} = {\thf} \left(\frac{m_1}{m_R}\right)_{ej} 
\omega (\xi+3)\tan^2\theta.  
\ee
In standard renormalisations of GSW theory [7,31], the right 
and left components of the electron propagator carry different factors 
$Z_{2L}$, $Z_{2R}$. 

The up and down quark propagators are renormalised in the same way. To one loop 
we find that 
\begin{eqnarray}
a_{uj} &=& {\toe} \omega \xi [3 +\tfrac{17}{9} t^2 +(3-\tfrac53 t^2)\gamma^5], \quad
b_{uj} = \tfrac19 \omega (\xi+3) t^2 m_{uj1}, \\
a_{dj} &=&  {\toe} \omega \xi [3 +\tfrac{5}{9} t^2 +(3-\tfrac13 t^2)\gamma^5], \quad
b_{dj} = -\tfrac1{18} \omega (\xi+3) t^2 m_{dj1}. 
\end{eqnarray}
The condition (86) then requires that
\be
\kappa_{uj5} = a_{uj5}, \qquad \kappa_{dj5} = a_{dj5},  
\ee
leading to two expressions for $c_{2qjL}$:
\begin{eqnarray}
c_{2qjL} &=& \tfrac14 \omega \xi (3+\tfrac19) t^2 - \tfrac19 
\left({\frac{m_1}{m_R}}\right)_{uj} \omega  (\xi+3)t^2   \\ 
&=& \tfrac14 \omega \xi (3+\tfrac19) t^2 + {\tfrac1{18}} 
\left({\frac{m_1}{m_R}}\right)_{dj} \omega  (\xi+3)t^2   
\end{eqnarray}
so that 
\be
\left(\frac{m_1}{m_R}\right)_{dj} = -2\left(\frac{m_1}{m_R}\right)_{uj}. 
\ee
Then we obtain a common renormalisation factor for the up, down quarks of 
generation $j$, which to $O(g^2)$ is
\be
\widehat Z_{2qj} = 1-\Delta_{qj}, \quad\quad
\Delta_{qj} = {\ton}\left(\frac{m_1}{m_R}\right)_{uj} \omega  (\xi+3)\tan^2\theta.  
\ee
In addition, we obtain
\begin{eqnarray}
c_{2ujR} &=& {\tfn}\omega \xi t^2 - {\ton}\left(\frac{m_1}{m_R}\right)_{uj} \omega (\xi+3) 
t^2 , \\
c_{2djR} &=& {\ton}\omega \xi t^2 + \tfrac1{18}\left({\frac{m_1}{m_R}}\right)_{dj} 
\omega (\xi+3) t^2 . 
\end{eqnarray}

In the absence of confinement with only electroweak interactions present, a 
stable quark, like a stable lepton, can appear as an isolated particle in 
{\it (in,out)} states in our perturbative formalism. Assuming that the lightest 
quark 
is the (first generation) $u$ quark, $u$ is stable and the $S$-matrix reduction 
formula requires that $m_{u11}= m_{u1R}$. Then (107) gives
\be
m_{d11} = -2 m_{d1R}  
\ee
for the $d$ quark, so that $d$ must be unstable in this theory. (The 
quantisation procedure does not require that $m_{d11}>0$.) This result is not 
in conflict with the instability of $d$ seen in neutron decay. Equation (107) 
also shows that at least one of  $c,s$ and one of $t,b$ must be unstable in 
this theory; of course, physically they all are. 

\bigskip\bigskip
\noindent{\large\bf 5. Vertices}
\medskip

In this section we outline a way to 
renormalise vertices to all orders, using the $WWZ$ vertex as an example. Then we 
complete the renormalisation of the whole theory to one loop. 

For the $WW$Z vertex, expressed in terms of $W^+$, $W^-$, $Z$ lines carrying 
incoming momenta and indices $p,\lambda$; $q,\mu$; $r,\nu$ respectively, the 
unrenormalised coupling is
\begin{eqnarray}
v_{WWZ} &=& -ig\cos\theta [(r-q)_\lambda g_{\mu\nu} +(q-p)_\nu 
g_{\lambda\mu}+(p-r)_\mu g_{\nu\lambda} ] \nono \\
&=& g f_{WWZ} .  
\end{eqnarray}
All the one-loop and counterterm contributions contain the factor $v_{WWZ}$, and 
we shall assume that this holds true to all orders. Renormalising, and 
factoring out $f_{WWZ}$, we obtain the vertex quantity
\begin{eqnarray}
G_R &=& g(1+d_1-c_1) \widehat Z_W \widehat Z_Z^{1/2} \nono \\ 
&=& [1+O(\epsilon)] G_{RL}    
\end{eqnarray}
where $d_1$, $c_1$ are the vertex part and counterterm and 
\be
G_{RL} = g(1+d_{1L} -c_1) \widehat Z_W Z_Z^{1/2}, 
\ee
in which $\widehat Z_W$, $\widehat Z_Z$ are given by (36) and
\be
d_{1L} = \sum d_{1(2n,n)} g^{2n} \epsilon^{-n}  
\ee
is the leading divergent part of $d_1$ (c.f.\ $\pi_{NL}$). We have anticipated 
that $c_1$ is a series in $g^2\epsilon^{-1}$. 

With $d_{1L}$, $c_1$, $\hat\pi_{WL}$, $\hat\pi_{ZL}$ correct to $O(g^{2N})$, 
$N$ arbitrary, $G_{RL}$ is then given correctly to $O(g^{2N})$ by
\begin{eqnarray}
G_{RL}
&=& g(1+d_{1L} -c_1) (1-\hat\pi_{WL} +\hat\pi^2_{WL}-\cdots) (1-
{\thf}\hat\pi_{ZL} +\tfrac3{8}\hat\pi^2_{ZL} - \cdots) \nono \\
&=& g\left[1+\sum(d_1-c_1)_{(2n,n)} g^{2n} \epsilon^{-n} \right] \left[ 1+\sum P_{2n} g^{2n} 
\epsilon^{-n} \right] , 
\end{eqnarray}
where
\begin{eqnarray}
P_2 &=& - (\hat \pi_{WL2} +{\thf} \hat\pi_{ZL2}),  \\ 
P_4 &=& - (\hat\pi_{WL4} + {\thf} \hat\pi _{ZL4}) +\hat\pi^2_{WL2} + \tfrac38 
\hat\pi^2_{ZL2} +{\thf} \hat\pi_{WL2}\hat\pi_{ZL2},  
\end{eqnarray}
and so on. 
It is clear that we can choose $c_1$, order by order,
\begin{eqnarray}
c_{1(2)} &=& d_{1(2,1)} +P_2 , \\  
c_{1(4)} &=& d_{1(4,2)} + [d_{1(2,1)} - c_{1(2)}] P_2 +P_4,  
\end{eqnarray}
and so on (obtaining $c_1 = \sum c_{1(2n)} g^{2n} \epsilon^{-n}$) such that 
$G_{RL} = g$ to $O(g^{2N})$, $N$ arbitrary. Then, from (113), we have, in the 
limit $\epsilon \to 0$, 
\be
G_R = g  
\ee
to all orders. We have renormalised the $WWZ$ vertex to obtain the coupling 
parameter $g_R$, with $g_R =g$; and with $\theta$, in $f_{WWZ}$, unchanged, 
i.e.\ we have $\theta_R=\theta$.

We now complete the renormalisation of the whole theory to one loop. We take 
$m_1 = m_R$ for every charged lepton and up quark (we recall that this does not 
restrict the imaginary part of the mass of an unstable fermion). We obtain 
the standard coupling at each vertex, with $g_R = g$, $\theta_R=\theta$. It is 
convenient to drop unnecessary suffixes in what follows.

An analysis of loops shows that all boson--boson and boson--ghost vertices 
renormalise with $g_R = g$, $\theta_R=\theta$ provided that the equations
\begin{eqnarray}
d_1 - c_1 -\tfrac32\hat\pi &=& 0,  \\ 
d_5 - c_5 -2\hat \pi &=& 0,      \\
\tilde d_1 - \tilde c_1 -\tilde \pi_0 +\tilde c_3 -{\thf} \hat\pi &=& 0, 
\end{eqnarray}
hold, where $\hat\pi$ is given by (63), $\tilde\pi_0$ by (60), and $d_1$, 
$d_5$, $\tilde d_1$ are reduced three-boson, four-boson and boson--ghost vertex 
parts, from which certain coupling constant and momentum-dependent terms have 
been factored out. By calculation and using $SU(N)$ results [28] we obtain
\begin{eqnarray}
d_1 &=& {\tos} \omega  (9\xi - 17 +8n), \\  
d_5 &=& {\tot} \omega  (6\xi -4 +4n), \\ 
\tilde d_1 &=& \omega \xi.  
\end{eqnarray}
Then (120), (121), (122) give
\begin{eqnarray}
c_1 &=& {\tos} \omega  (9\xi - 17 +8n) - \tfrac32 \hat\pi, \\ 
c_5 &=& {\tot}\omega  (6\xi -4 +4n) - 2\hat\pi, \\ 
\tilde c_3 - \tilde c_1 &=& -{\thf} \omega (\xi +3) +{\thf} \hat\pi.   
\end{eqnarray}
The values of $\tilde c_3$, $\tilde c_1$ separately are not determined by 
renormalisation to one loop. 

The renormalisation to one loop of the lepton--boson vertices is similarly 
straightforward. For the $e(\mu,\tau)\nu W$ vertices to renormalise to the 
standard coupling, we must have
\be
\frac{-ig}{2\sqrt2} (1-\gamma^5) = \frac{-ig}{2\sqrt2} (1-\gamma^5) [1-
\Delta_\ell - {\thf}\hat\pi -c_{1\ell L} +{\tqr}(1-\tan^2\theta) v_1 -v_2],  
\ee
where the terms $v_1 = -\omega \xi$, $v_2 = -\tfrac32 \omega (\xi+1)$ come from the vertex 
parts comprising one boson and two fermion lines, and two boson and one fermion 
lines respectively. This gives, using (101) with $m_1 = m_R$,
\be
c_{1\ell L} = {\tqr} \omega [\xi(5-\tan^2\theta) +6(1-\tan^2\theta)] -{\thf}\hat\pi .  
\ee
The zero $\nu\nu A$ vertices require that
\begin{eqnarray}
c_{\ell L} &=& {\thf} g \tan\theta (c_{1\ell L} - v_1 +v_2) \nono \\
&=& g\tan\theta \{ {\toe} \omega  [3\xi -(\xi+6)\tan^2\theta] - {\tqr} \hat\pi \},  
\end{eqnarray}
and these values of $c_{1\ell L}$, $c_{\ell L}$, independent of $j$ 
(generation), renormalise the $\nu \nu Z$ vertices correctly. Then the $eeZ$ 
(etc.) and $eeA$ (etc.) vertices are renormalised correctly on choosing
\be
c_{eR} = {\thf} g\tan\theta [\omega (\xi - 3) \tan^2\theta -\hat\pi].   
\ee

The renormalisation of quark vertices goes through similarly. We obtain, taking 
$m_1 = m_R$ for each up quark, the generation-independent results that
\begin{eqnarray}
c_{1qL} &=& {\tqr} (1-{\ton} \tan^2\theta) v_1 -v_2 -\Delta_q -{\thf}\hat\pi ,\nono \\
&=& {\tqr}\omega  [5\xi +6 -{\tot}(\xi+4) \tan^2\theta] - {\thf}\hat\pi,  \\ 
c_{qL} &=& g\tan\theta \{ \tfrac1{12}\hat\pi - {\toe}\omega  [\xi(1-
{\ton}\tan^2\theta) - \tfrac49\tan^2\theta]\} , \\ 
c_{uR} &=& g\tan\theta \{{\tot}\hat\pi - \tfrac29 \omega  (\xi-1) \tan^2\theta \}, \\  
c_{dR} &=& g\tan\theta [-{\tos}\hat\pi - {\ton} \omega \tan^2\theta]. 
\end{eqnarray}

To one loop the $SU(2)_L\times U(1)_Y$-invariance conditions (7) are
\begin{eqnarray}
c_W -c_1 &=& c_1 - c_5 \nono\\
&=& \tilde c_3 -\tilde c_1 \nono\\
&=& c_{2\ell jL} - c_{1\ell jL} \nono\\
&=& c_{2qjL} - c_{1qjL},  
\end{eqnarray}
where, with $m_1=m_R$ for the charged leptons and up quarks, the $c_{..jL}$ are 
independent of the generation index $j$. We see that these equations are 
satisfied by the values given by (61), (128), (129), (130), (96), (132), (105) 
and (135). 

The renormalisation of the theory to one loop is complete, with the condition 
(7) satisfied, and with the bosons and fermions possessing appropriate real and 
complex masses.

Renormalised to one loop, the $W$, $Z$, photon, ghost and fermion propagators, 
and the coupling parameters $g$, $\theta$, are independent of the gauge 
parameter $\xi$. Then the resulting $S$-matrix elements are independent of 
$\xi$.

\bigskip\bigskip
\noindent{\large\bf 6. Conclusions and discussion}
\medskip

We have set up a perturbative $SU(2)_L\times U(1)_Y$ electroweak theory 
containing $W$, $Z$, photon, ghost, lepton and quark fields, but no Higgs 
field. We make an unconventional choice for the unperturbed Lagrangian 
$\call_0$. Then physical $W$, $Z$ and fermion masses are obtained on 
renormalisation to one loop by a somewhat unorthodox but systematic method. The 
renormalisation preserves the usual couplings, with renormalised parameters 
$g_R$, $\theta_R$ equal to the original parameters $g$, $\theta$, requires 
neutrinos to be massless and $m_W = m_Z\cos\theta$ to hold, and causes the 
$A\leftrightarrow Z$ mixing to drop out. Also, the theory 
renormalised to one loop gives $S$-matrix elements that are independent of the 
gauge parameter $\xi$.

The theory should be renormalisable to all orders, as we have outlined such a
systematic procedure to renormalise the boson and fermion propagators and 
the vertices, to arbitrary order. As 
discussed in section~1, a treatment of the unitarity of this theory, which 
contains unstable particles (and similarly a proper treatment of unitarity in 
the Standard Model), waits for an extension of perturbation 
theory that incorporates unstable particles consistently. 

The choice of Dirac, not Majorana, masses~(2) for the fermions forced the 
neutrinos to be massless, as neutrinos lack right-handed partners in the 
Standard Model. In turn, this choice implied a charge-neutral vacuum, unbroken
lepton number and QED gauge symmetries, and a massless photon~(55). Furthermore,
careful consideration of the $SU(2)_L$ symmetry properties of the $\hat\pi_J$
shows that the phenomenologically necessary relation~(62) between $m_W$ and 
$m_Z$ is automatic. Majorana fermion masses would violate this requirement,
but they should be explored as a theoretical possibility.

The mechanism of fermion and gauge boson mass generation (dynamically broken 
chiral and gauge symmetries) presented here and earlier [3] bears similarities
to existing models, but with some important differences. The fermions
breaking chiral and gauge symmetries are the Standard Model fermions themselves,
with no new particles or interactions necessary, unlike technicolour [32] and 
top condensate [33] models. And our scheme is perturbative in the gauge
coupling. This feature differs from the non-perturbative ``soft'' fermion 
self-energy ansatz for dynamically breaking chiral gauge symmetries discovered 
by Jackiw and Johnson [34] and applied to the Standard Model by Carpenter {\it
et al.} [35]. In that scenario, the gauge boson and fermion mass(es) are
linked by a finite relationship {\it after} renormalisation that predicts too
large a top quark mass (or equivalently, too small $W$ and $Z$ masses). Our
scheme renormalises the fermion and gauge boson masses separately, so that
their masses remain {\it independent.}

On the other hand, simple Ward identity arguments [36] imply in all of these 
models an effective Higgs-like scalar resonance with mass $m_H\simeq\sqrt{2}
m_t\simeq$ 250 GeV, independent of model solution details. The question of 
choosing
the perturbative or non-perturbative solution can only be settled by explicit
calculation of the vacuum energy, an issue, like unitarity and Majorana masses, 
we leave to a subsequent treatment.

\bigskip\bigskip\newpage
\noindent{\large\bf Acknowledgments and Note}
\medskip

D.C.K. thanks V.~A.~Miransky of the Bogolyubov Institute for Theoretical 
Physics, Kiev, for helpful discussions and Betty Nicholson and Ann Milligan 
for their help with completing the manuscript.

This paper is dedicated to the memory of Angus Nicholson (1927-1999).  A native
of Australia and receiving his M.Sc. and Ph.D. degrees in physics at London
and Birmingham Universities, he spent much of his working life as research 
scientist with the Australian Defence Science and Technology Organisation.  
After his retirement, he continued theoretical physics work as visiting scholar 
at The Australian National University, Canberra, until his death.

\end{document}